\documentclass[iop]{emulateapj} % read by: ACROREAD; emulateapj-rtx4
\shortauthors{Sekanina}
\shorttitle{Orbital Period and Pedigree of a Sungrazing Comet}
\slugcomment{Version \today }

\begin{document}
%
% \title{Why Knowledge of the Orbital Periods of Sungrazing Comets\\Is
% Vital to Advanced Models of the Kreutz System}
% Carefully Determined
%
\title{Orbital-Period Determination As an Indispensable Tool to Study\\the
 Pedigree of a Sungrazing Comet}
\author{Zdenek Sekanina}
\affil{La Canada Flintridge, California 91011, U.S.A.; {\sl ZdenSek@gmail.com}}

\begin{abstract} % maximum length = 1920 characters
Among some 4500 Kreutz sungrazers known, the orbital period has been
established~to~\mbox{better}~than about $\pm$20~years only for C/1882~R1,
C/1963~R1, C/1965~S1, C/2011~W3,~and~C/2026~A1.  I~\mbox{describe}
solutions to a range of intriguing problems involving the
orbital-period~\mbox{determination}.~A~\mbox{helpful},~but
computer-intensive routine is a detailed investigation of
derived orbital periods as a function~of~the~last
observation used, which allows one to filter out effects of activity (the
case of C/2026~A1)~or~\mbox{nuclear} fragmentation (the case of C/1965~S1)
and thereby reliably evaluate the time of the previous perihelion, a
cornerstone in the quest for a sungrazer's pedigree.  Experience shows that
only very massive~objects, such as the original nucleus of C/1882~R1 or its
main fragment~B are immune to effects of this kind.  Different problems
are presented by a sungrazer whose nucleus falls apart shortly after
perihelion (the case of C/2011~W3) or by one observed only after perihelion
(the case of C/1963~R1).  The documented high sensitivity of the derived
orbital period to minor perturbations of a sungrazing comet's motion~is
exploited to advantage when the standard approach does not work or
yields inconclusive results.
\end{abstract}
\keywords{comets:\ C/1882\,R1, C/1963\,R1, C/1965\,S1, C/2011\,W3,
 C/2026\,A1; methods:\ data analysis\vspace{-0.1cm}}
\section{Introduction} % Section 1
%
%\vspace{-0.04cm}
Of the approximately 4500 Kreutz sungrazing comets known at
present, the orbital period has been determined from observations
to better than about $\pm$20~years for not more than five.
Chronologically they are:\ C/1882~R1, C/1963~R1, C/1965~S1,
C/2011~W3, and C/2026~A1.~In this paper I describe some unexpected
findings that computation of the orbital period was revealing
for three of these objects:\ comet MAPS of 2026, a dwarf sungrazer;
Ikeya-Seki of 1965, the brightest comet of the 20th century; and
the Great September Comet of 1882,~a~magnifi\-cent spectacle.
Less elaborate conclusions are drawn on the orbital periods of the
two remaining sungrazers.
% Lovejoy of 2011, the penultimate.

Unlike for unrelated comets, knowledge of the orbital period 
of a Kreutz sungrazer not only expands our understanding of
the object itself but likewise contributes to an improved
awareness of the Kreutz system as a whole because of the
relationships among its members as fragments of a single
progenitor.  The spatial and temporal distributions of the
Kreutz comets are products of an enormous number of events
of cascading fragmentation, depending in part on the orbital
periods of parent bodies.  The orbital period of a sungrazer
is employed~to~track members of the previous generations, as
demonstrated~in detail below, and proves helpful in an effort
to chart the object's ancestry. 

Apart from those issues, the orbital periods of Kreutz comets
are shown to be a potential tool for sorting out effects of
outgassing-driven nongravitational forces on the sungrazers'
motions, thereby providing an avenue for more accurately
determining the time of an object's previous passage through
perihelion.

Finally, since the magnitude of nongravitational effects
is object-size dependent, generally increasing as the size
decreases, the extent of variations in the orbital period
of a Kreutz comet also offers ballpark information on the
dimensions of its nucleus.

\section{Comet C/2026 A1 (MAPS)}   % Section 2
Investigated by Sekanina \& Kr\'olikowska (2026), this comet
displayed two intriguing traits:\ (i)~it had an exceptionally
long orbital period, exceeding 1600~years, about twice an
average of the remaining four objects; and (ii)~it apparently
was accompanied by at least three, but possibly many more,
distant dwarf sungrazers. 

These companions offered circumstantial evidence that at the time of
separation from the massive parent body at the previous perihelion,
C/2026~A1 apparently was~part of a somewhat larger object that much
later fragmented, perhaps after having passed aphelion.  Affecting
the~\mbox{orbit} of comet C/2026~A1, the breakup and birth of the
companions were in fact a warning (that we were of course unaware
of) of a high degree of friability of its nucleus, in the light of
which the comet's disintegration, completed about two hours before
perihelion (Sekanina 2026a), was hardly surprising.

The anomalously long orbital period was a product of
the conditions at the time of separation from the
parent sungrazer (Sekanina 2026b) and is understood
in the context of the contact-binary hypothesis
(Sekanina~2021).  Unlike the other sungrazers with known
orbital periods, the object, whose largest fragment
arrived as C/2026~A1, did not break off from its parent
in the 11th or 12th~century, but in AD~363, when the
first-generation fragments of the progenitor appeared
following the birth of the Kreutz system.  As part of
an {\small \bf outlying} piece of one in a group of
spectacular daylight comets observed that year, C/2026~A1
became the only known second-generation fragment of
Aristotle's comet to arrive nine to ten centuries later
than one would expect.  To end up in the observed orbit,
its distance from the massive parent's center must have been
more than $\sim$12~km (Sekanina 2026b) and the parent
must have been at least $\sim$25~km across, in agreement
with other constraints on the dimensions of the daylight
comets of AD~363 (Sekanina 2021, 2022).

The results of Sekanina \& Kr\'olikowska's (2026)
computations confirmed the suspicion that the derived
orbital period of comet C/2026~A1 depended on the time
of the last ground-based observation used to determine
the orbit.  As the comet was observed from the ground down
to a heliocentric distance of $\sim$0.4~AU, Sekanina \&
Kr\'olikowska suggest that their results provide evidence on
effects of increasing {\small \bf erratic} activity on the
comet's motion:\ the closer to perihelion the employed arc of
the orbit extended, the longer was the derived orbital~\mbox{period},
a longer period implying an earlier time of the~\mbox{previous}
perihelion.  The orbital solutions based on observations
more than \mbox{50--53}~days before perihelion, when the
heliocentric distance of the comet exceeded $\sim$1.55\,AU, were
the only ones that gave a true orbital period compatible with the
expected perihelion time in AD~363 to within 1$\sigma$.~Sekanina
\& Kr\'olikowska pointed out that \mbox{50--53}~days before
perihelion happened to be the time of a major anomalous feature
in the light curve, when the steep increase in the comet's
intrinsic brightness abruptly ceased, followed by a much slower
and noisier climb.  An emission source on the nucleus may have
gradually become deactivated.  The relevant part of the comet's
light curve is reproduced in Figure~1.

Sekanina \& Kr\'olikowska's results revealed that the derived
orbital period was highly responsive to minor deviations from
the gravitational law.  This result~suggests
that~the~sizable~progressive~orbital-period~variations~of
C/2026~A1, resulting from fitting astrometric observations over
various arcs of the orbit, could serve as a delicate sensor of
otherwise undetectable nongravitational effects in its motion.
This explains the increasing deviation from the genuine orbital
period of 1663~years with decreasing heliocentric distance, as
the comet's pattern of outgassing grew ever more complex and
consequential.\,
%
% extraordinarily high responsiveness of this procedure explains
% an apparent contradiction, as a standard, purely gravitational
% orbital fit left positional residuals whose distributions appeared,
% by visual inspection, quite acceptable in both right ascension
% and declination over the entire observed orbital arc, up to
% seven days before perihelion.

%
\begin{figure}[hb] % Figure 1
\vspace{0.45cm}
\hspace{-0.21cm}
\centerline{
\scalebox{0.65}{
\includegraphics{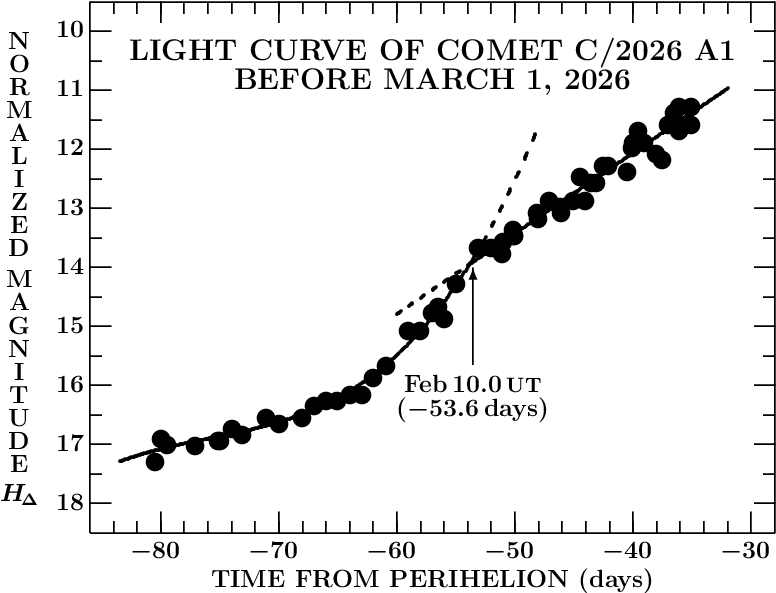}}}
\vspace{-0.03cm}
\caption{Light curve of comet C/2026 A1 between January~14
and February~28, 2026, or 80 to 35~days before perihelion.
The observed total CCD~magnitude is normalized to unit
geocentric distance and a zero phase angle.  Note the
deflection point on February~9--12, when the early rate of
rapid brightening abruptly changes to a much slower and
noisier climb.  The anomaly may~sig\-nal an onset of deactivation
of a gas-emission source on the comet's nucleus.  The data
come from the {\it Comet Observation Database\/} (COBS)
website maintained by the \v{C}rni Vrh Observatory,
Slovenia. (Reproduced from Sekanina \& Kr\'olikowska
2026.){\vspace{0.16cm}}}
\end{figure}

\section{Comet C/1965 S1 (Ikeya-Seki)} % Section 3
The brightest comet of the 20th century, this sungrazer was
a major source of motivation for Marsden's~(1967) pioneering
paper on the sungrazing comets.  The object was discovered
a little more than one month before perihelion and observed
astrometrically over a period of nearly four months.  The
nucleus split tidally at perihelion, which was a complication
that led to the appearance of a pair of nuclear fragments
starting two weeks later.  Marsden derived separate sets
of orbital elements for either mass by linking observations
of the single nucleus before perihelion with those of the
respective nuclear component afterwards.  Although this
obviously is dynamically questionable, it is a virtually
standard procedure applied to the split comets.  Its aim
is to improve the determinacy of the orbital solution.  In
the case of comet Ikeya-Seki, the linkage left no systematic
trends in the distribution of positional residuals for either
mass.  The osculating orbital periods for the brighter and
fainter nuclei were near 880~years and 1055~years, respectively.

The orbital elements of comet Ikeya-Seki were strikingly similar
to those of the Great September Comet of 1882 (C/1882~R1; see
Section~4).  Their angular elements differed by less than
0$^\circ\!$.7 and their perihelion distances by less than
0.008$\:R_\odot$ (or 5500~km).  This correspondence gave rise
to two connected questions, both scrutinized by Marsden:\
(i)~Could the two sungrazers be fragments of a common parent that
they separated from at their previous perihelion?  And, (ii)~could
the famous Great Comet of 1106 (X/1106~C1) be identified with
that parent?  In practice, Marsden answered the second question
in two parts, before addressing the first question.

Long before Marsden, the second issue was tackled, to a degree,
by Kreutz (1888), whose computations led to a conclusion that
the daylight observations of the Great Comet of 1106 supported
the notion of its identity with the Great September Comet of
1882.  While Marsden did tend to agree with Kreutz on this
issue, he made a concerted effort to thoroughly investigate this
potential~rela\-tionship.  And he began his exercise with comet
Ikeya-Seki.

Integrating his orbit for the brighter component of the
nucleus, Marsden computed that in the baseline case the previous
perihelion took place in September 1116,~about ten
years after the arrival~of~the~Great~Comet, and~concluded
that, given the hidden uncertainties in the
comet's orbital elements, the date was in fact ``{\it
remarkably close\/}'' to the time of appearance of the comet
of 1106 and that ``{\it of all the comets reported
during the eleventh and twelfth centuries, this seems by far
the most promising candidate for the previous appearance of
comet\/} [{\it C/1965~S1\/}].''

Marsden's next step was integration back in time of the
motion of the principal nucleus B of the Great September
Comet of 1882.  He used Kreutz's (1891) elements III to
begin with and determined that the previous perihelion
occurred in April 1138.  Although this outcome looked like
a setback to a hypothesis proclaiming identity between
the comets of 1882 and 1106, Marsden argued that because
the observations used by Kreutz were visual (and therefore
less accurate than the photographic observations of comet
Ikeya-Seki) and the nuclear region of the 1882 comet had
a complex appearance, the poorer agreement was to be
expected.

On the other hand, it should be pointed out that
the overall span of the orbital arc was 260~days with
about 1500~observations of the principal nucleus of
the 1882 comet, but only 115~days with 119~observations
of the principal nucleus of comet Ikeya-Seki.  In
addition, the dimensions of the principal nucleus of
the 1882 comet must have been significantly
larger than the dimensions of the principal nucleus of
comet Ikeya-Seki, so that the errors caused by neglect of
the nongravitational effects in the motion of the latter
are expected to exceed those in the motion of the former.
As a rule, neglect of these effects increases the orbital
period and leads to a date of the previous perihelion
that is {\small \bf earlier} than the actual date.

Given these arguments and the discrepancy of 32~years
between the arrival time of the 1106 comet and the
predicted time of the previous perihelion of the 1882
comet,~I felt that the identity of the two objects was
questionable and thought that it was highly desirable
to have~another look at the whole issue.

Having investigated separately the evolution of the
orbital motions of comet Ikeya-Seki and the Great
September Comet of 1882, Marsden was in a position
to establish with high probability whether they were
fragments of the same parent.   As it affects the
outcome only weakly, the choice of the common
perihelion time to which the motions of the two
sungrazers were to be integrated was not critical.
Marsden chose the year 1115 and %\vspace{-0.01cm}
found that their angular elements agreed within
0$^\circ\!$.02 and their~peri\-helion distances
within 0.0013$\:R_\odot$ or 900~km. %\vspace{-0.01cm}
Their separation from each other in the 12th century
was thereby with near certainty confirmed.

Following the formulation of a new, contact-binary~hypothesis
for the Kreutz system (Sekanina 2021), it was
desirable to verify this model's features by
integrating the orbital motions of the main
members.  I tackled the task in collaboration with
R.~Kracht (Sekanina \& Kracht 2022) and we included
comet Ikeya-Seki in order to examine its presumed
relationship with the Great Comet of 1106.  First,
we were able to closely reproduce Marsden's (1967)
relativistic orbit for Ikeya-Seki, with the orbital
period differing from his result by merely one year
(or 0.11~percent), and with the time of the previous
perihelion predicted for February~1116.

In the next phase of our investigation of comet Ikeya-Seki
we computed two nongravitational orbital solutions, which
Marsden skipped.  When we solved for just the radial
component of the nongravitational acceleration, the
orbital period changed hardly at all, but by solving for
both the radial and transverse components we obtained a
dramatically different outcome:\ the previous perihelion
time moved by 36~years to May 1152!  Since nongravitational
solutions are known to be sometimes misleading, we took
this result with a grain of salt, even though the mean
residual dropped from $\pm$1$^{\prime\prime}\!$.47 to
$\pm$1$^{\prime\prime}\!$.37.

Because the time of the previous perihelion now differed
from the arrival time of the Great Comet of 1106 so
dramatically, we applied a new method of attacking
the problem by employing a more conservative, gradual
approach.  The starting point was the orbit that we
already computed; it was derived from the combined
observed arcs of the single nucleus (from September~21
through the perihelion breakup) and the principal
post-perihelion nucleus (from the breakup through
January~14, 1966).  The preperihelion arc was 30~days
long, the post-perihelion arc 85~days.  We contemplated:\
if the motion of this principal part of the fragmented nucleus
was \mbox{\small \bf measurably}~affected by the breakup
at perihelion, the {\small \bf impact} of the breakup
on the nominal orbit (and the date of the
previous perihelion) {\small \bf should diminish,
when the orbital length of the nuclear fragment
contributing to the nominal orbit is curtailed}.
Specifically, if the nuclear fragment ended up in
an orbit with a longer period than was the period
of the single nucleus before its breakup, a shorter
arc of the nuclear fragment should result in a
shorter period of the nominal orbit, and the comet's
previous perihelion should have occurred later, and
vice versa.  If the Great Comet of 1106 was indeed
the parent of Ikeya-Seki, a lesser impact of the
motion of the nuclear fragment should result
in an {\small \bf earlier} date of the previous
perihelion.  A {\small \bf later} date should rule out
the comet of 1106 as the parent.

By eliminating the observations of January~14, 1966,
the observed arc extended only to December~31, 1965,
71~days after breakup, down from 85~days.  The new
nominal orbit suggested that the previous perihelion
took place in November 1118, 2.8~years {\small \bf
later}.  It looked as though comet Ikeya-Seki {\small
\bf was not} a fragment of the Great Comet of 1106.
However, it was desirable to rule out possible noise
in the positional data, and we continued to gradually
shorten the orbital arc of the nuclear fragment.
Reducing the arc length from 71~days to 64~days changed
the predicted time of the previous perihelion to the
beginning of December 1119; reducing it to 47~days
gave May 1124; etc.\ until reducing it to 16~days after
breakup gave August 1136.  Because the two fragments
of the split nucleus were not optically resolved until
November~4 (Pohn 1965), or 14~days after perihelion, the
orbital arcs ending before November~4 could not be combined
with the arcs ending after this date.  The overall
conclusion is absolutely clear:\ The Great Comet of
1106 could not be the parent of comet Ikeya-Seki and,
by extension, of the Great September Comet of 1882
either --- an important pedigree result.

\begin{figure*}[t] % Figure 2
\vspace{0.2cm}
\hspace{-0.28cm}
\centerline{
\scalebox{0.6}{ % 0.825
\includegraphics{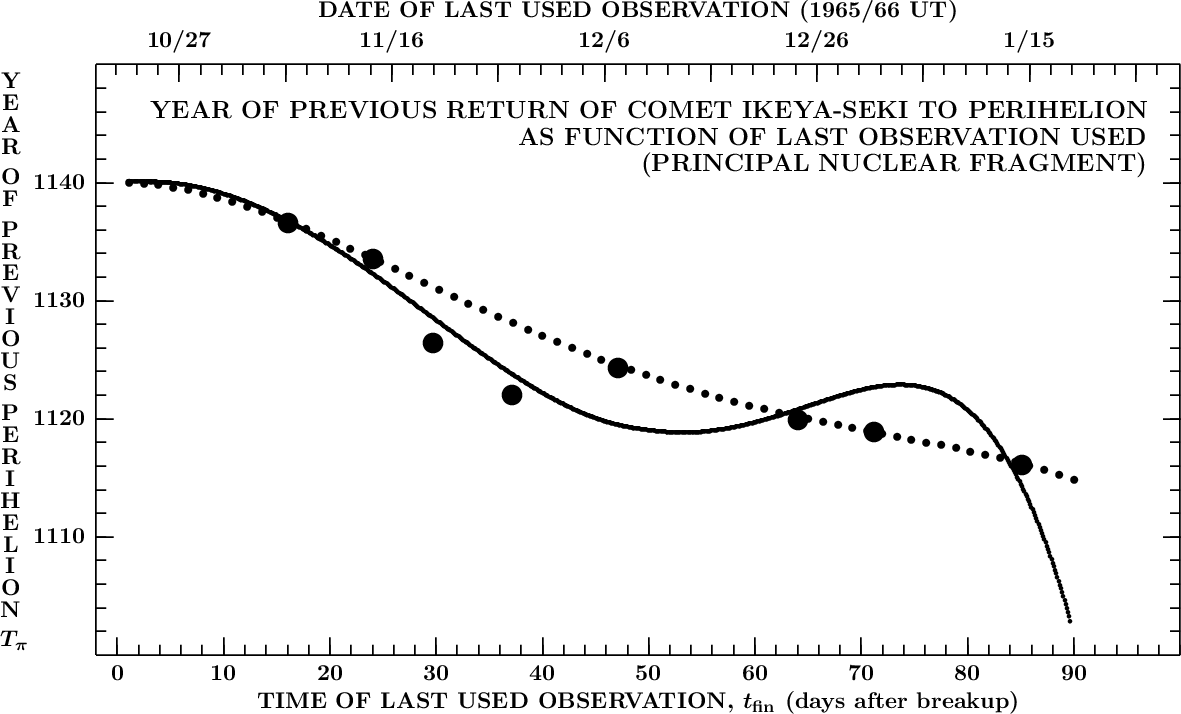}}}
\caption{Year of the previous perihelion passage of comet
 Ikeya-Seki in the 12th century, computed from the comet's
 orbit in 1965 (observations of the single nucleus before
 perihelion and the principal mass of the split nucleus
 after perihelion) truncated by different last observations
 used, reckoned from the perihelion breakup.  The first
 observation was always on Sept 21, 1965, one month
 before perihelion.  The disks are the results from the
 orbits ending on the dates (from right to left):\ Jan~14,
 1966; Dec~31, 24, and 7, 1965; Nov~27, 19, 14, and 6,
 1965.  The plot shows that the computed date of the
 previous perihelion was gradually shifting to later
 years as the orbital arc was getting shorter and the
 contribution from the principal post-perihelion nucleus,
 which ended up in an orbit of a longer period than was
 the period of the single nucleus, was diminishing.
 Extrapolation to the time of breakup $(t_{\rm fin} = 0)$
 yiels the predicted time of the previous perihelion in
 1140.  The solid curve represents the cubic polynomial
 fit to function ${\cal F}$, expressing $T_\pi$ by
 Equation~(5); the dots show an alternative quadratic
 solution that fits exceptionally well six of the eight
 data points. (Expanded from Sekanina \& Kracht
 2022.){\vspace{0.5cm}}}
\end{figure*}

To extend this exercise, I now search for the time of
the previous perihelion of comet Ikeya-Seki, $T_0$,
using the dates in the 12th century derived from the
orbits truncated to between 16.02~days and 85.15~days
after the 1965 perihelion, as described above.  Let
$T_\pi(t_{\rm fin})$ be the date of the comet's
previous perihelion, determined from an orbit truncated
at time $t_{\rm fin}$, as shown in Figure~2.  Because
$t_{\rm fin}$ is reckoned from the 1965 perihelion
(the time of breakup), time $T_0$ that I search
for is obtained by an appropriate extrapolation
to \mbox{$t_{\rm fin} \rightarrow 0$}, when indeed
\mbox{$T_\pi(t_{\rm fin}) \rightarrow T_0$}.  The
eight data points in the plot can be fitted with a
curve, approximated by a polynomial,
\begin{equation}
T_\pi(t_{\rm fin}) = T_0 + \sum_{k=1}^n \! c_k \:\!t_{\rm fin}^k,
\end{equation}
where $c_k$ are coefficients derived by the least-squares
fit and $n$ is limited by the number of data points.  In
order that $T_\pi$ smoothly approach $T_0$, one requires
that
\begin{equation}
\lim_{t_{\rm fin} \rightarrow 0} \frac{\partial
 T_\pi}{\partial t_{\rm fin}} = \!\lim_{t_{\rm fin}
 \rightarrow 0} \! \left( c_1 \!+\! 2\:\!c_2\:\!t_{\rm fin}
 \!+\! 3\:\!c_3\:\!t_{\rm fin}^2 \!+ \ldots \right) = 0,
\end{equation}
so that first of all,
\begin{equation}
c_1 = 0.
\end{equation}
Inserting it into Equation (1), one gets
\begin{equation}
{\cal F}(T_0,n;t_{\rm fin}) = \frac{T_\pi(t_{\rm
 fin}) \!-\! T_0}{t_{\rm fin}^2} = b_0 +
 \sum_{k=1}^n b_k\:\! t_{\rm fin}^k,
\end{equation}
where $b_k = c_{k+2} \, (k = 0, 1, 2, \ldots)$.
Function ${\cal F}(T_0,n;t_{\rm fin})$ was next
calculated for a variety of assumed values of
$T_0$ and the solution then optimized using the
mean residual as a function of exponent $n$.

Since the truncated orbital solutions indicated that
$T_0$ was most probably in the late 1130s, Equation~(4)
was solved for the beginning of each year in the period
of \mbox{$1131 \leq T_0 \leq 1145$}. Based on cursory
inspection of func\-tion ${\cal F}(T_0,n;t_{\rm fin})$, the
polynomials were limited to \mbox{$n > 2$}, as two of
the eight data points could not be fitted by a quadratic
function.  The upper limit of $n$ was constrained to
\mbox{$n \leq 6$} by the number of data points, $N$,
as I required that \mbox{$n \!+\! 1 \leq N \!-\! 1$}.
The polynomials with \mbox{$n = 4, 5,$ and 6} suggested
that the previous perihelion time was most probably in
1136 with the standard deviations of function $\cal F$
equaling, respectively, $\pm$0.00165, $\pm$0.00190, and
$\pm$0.00082~year per day$^2$.  The last solution was
in particularly good agreement with times $T_\pi$ derived
from the orbital computations, resulting in a standard
deviation of only $\pm$0.52~year.  However, since the
values of $T_\pi$ themselves were burdened by errors
exceeding one year, this agreement could be fortuitous.

An optimized solution with \mbox{$n = 3$} resulted in
\mbox{$T_0 = 1140$} and the following expression for $T_\pi$:
\begin{eqnarray}
T_\pi(t_{\rm fin}) & = & 1140 - 0.0058823\,t_{\rm fin}^2
 \left(1 + 0.12140\,t_{\rm fin} \rule{0mm}{3mm}
 \right. \nonumber \\[-0.05cm]
 & & \left. -0.0033553\,t_{\rm fin}^2 + 0.000022025\,t_{\rm
 fin}^3 \right).
\end{eqnarray}
The standard deviation for function ${\cal F}(1140,3;t_{\rm fin})$
was $\pm$0.00211~year per day$^2$ and the quality of fit to
the data is apparent from Figure~2.  The fit is clearly
not satisfactory near \mbox{$t_{\rm fin} \sim 80$ days}.
What counts, however, is that the expression in Equation~(5)
agress very well with the data at \mbox{$t_{\rm fin} < 40$ days}.

A peculiar feature of the plot in Figure 2 is that with the
exception of the data points near \mbox{$t_{\rm fin} = 30$ days}
and 37~days after perihelion, the other seem to be distributed
along an exceptionally smooth curve.  This impression is confirmed
by calculations, as function ${\cal F}(T_0,n;t_{\rm fin})$ for
the six remaining points is readily represented by a quadratic
polynomial.  The best solution has a very small standard
deviation of $\pm$0.000074~year per day$^2$, it does again
yield \mbox{$T_0 = 1140$}, and $T_\pi$ varies with $t_{\rm
fin}$ as follows:
\begin{equation}
T_\pi = 1140 \!-\! 0.01724\,t_{\rm fin}^2 \! \left(1 \!-\!
 0.01642\,t_{\rm fin} \!+\! 0.00008132\,t_{\rm fin}^2 \right)\!.
\end{equation}
The standard deviation of $T_\pi$ is merely $\pm$0.09~year, but
the data point near 30~days after perihelion requires a correction
of +4.7~years and the data point near 37~days needs a correction
of +6.0~years.  Formula~(6) is described by the dotted curve
in Figure~2.  Although very different, both solutions predict
\mbox{$T_0 = 1140$} and have another feature in common, as an
average rate of change in the time of the previous perihelion
over the observed post-perihelion arc of the orbit amounts in
either case to
\begin{equation}
\left\langle \! \frac{\partial T_\pi}{\partial t_{\rm fin}} \!
 \right\rangle = -0.28 \; {\rm year \; per \; day}.
\end{equation}

\begin{figure*}[t] % Figure 3
\vspace{0.2cm}
\hspace{-0.28cm}
\centerline{
\scalebox{0.7}{
\includegraphics{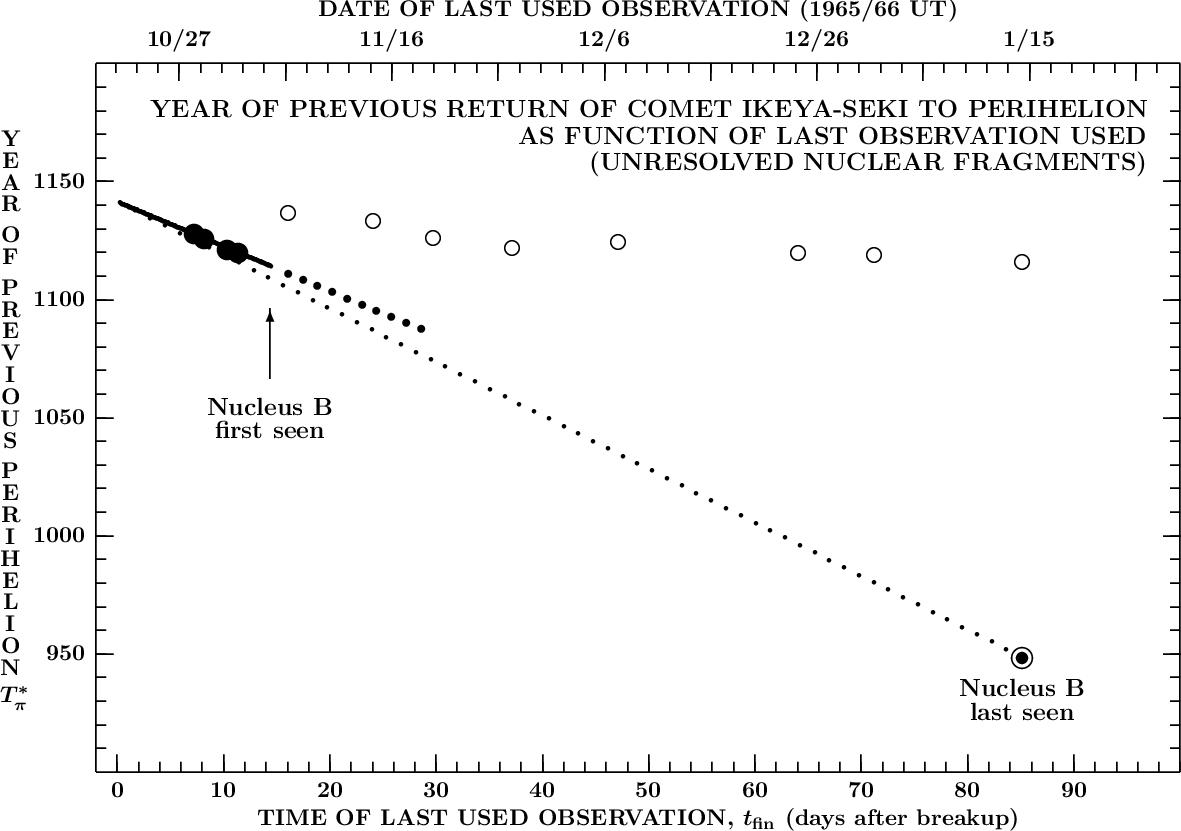}}}
\caption{Year of the previous perihelion passage of comet Ikeya-Seki
in the 12th century, computed from the comet's orbit in 1965
(observations of the single nucleus before perihelion and the
unresolved mass of the split nucleus after perihelion) truncated
by different last observations used, reckoned from the perihelion
breakup.  The first observation was always on Sept~21, 1965.
The disks are the results from the orbits ending on the dates
(from right to left):\ Nov~1, Oct~31, 29, and 28, 1965.  The
plot shows that the computed date of the previous perihelion was
rapidly shifting to later years as the orbital arc was getting
shorter.  Extrapolation to the time of breakup \mbox{$(t_{\rm fin}
= 0)$} yields the predicted time of the previous perihelion in 1141
(solid line).  The ordinate of nucleus~B at \mbox{$t_{\rm fin} = 85$
days} comes from its orbit by Marsden (1967).  The open circles
are the data points from Figure~2. (Expanded and modified from
Sekanina \& Kracht 2022.){\vspace{0.54cm}}}
\end{figure*}

Additional orbital solutions computed by Sekanina \& Kracht
(2022) used short orbital arcs, with the last observations on
or before November~2, 1965, less than 12~days after perihelion.
Although the two components of the fragmented nucleus were
predicted to be as far apart as 8$^{\prime\prime}$ on
November~1.5~UT, they still were optically unresolved.  As a
result, these solutions could not be combined with the
solutions for the principal fragment, computed from the
longer orbital arcs.
 
Although neither component of the fragmented nucleus of
comet Ikeya-Seki had existed before the 1965 perihelion,
the principal component is believed to have been the bearer
of much of the original mass of the object.  Efforts to
learn its past history were the primary driver behind
Marsden's (1967) linkage of the motion of that component
with the motion of the single nucleus before perihelion
and integration back to the 12th century.  However,
Marsden also linked successfully the motion of the
preperihelion nucleus with the fainter component B of the
post-perihelion fragmented nucleus, which demonstrated
that the orbit of the single nucleus was poorly defined.
Marsden did not integrate the linked motion of the fainter
nucleus back in time, and indeed this would make no sense.
But his osculating orbit for this fainter component,
combined with the known effect of the indirect planetary
perturbations on the principal nucleus, suggests that
conjecturally this nucleus would have previously passed
perihelion in the year 948, nearly two centuries earlier
than did the comet itself.  Ignoring that this inference
reflects primarily effects of a high nongravitational
acceleration, the divergence of nearly two centuries {\it
and\/} simultaneous compliance with the set of preperihelion
observations confirms that the orbital period of the
nucleus before fragmentation was indeterminate.

The orbital elements utilizing the astrometric data on
nucleus~B of Ikeya-Seki allow one to calculate for it
a hypothetical average rate of change in the time of
this fragment's ``previous perihelion,'' $T_\pi^\prime$:
\begin{equation}
\left\langle \! \frac{\partial T_\pi^\prime}{\partial t_{\rm
 fin}} \! \right\rangle = -2.25 \; {\rm years \; per \; day}.
\end{equation}

With the average rate established from the observations of
the principal component of the nucleus in condition (7),
$\langle \partial T_\pi^\prime/\partial t_{\rm fin} \rangle$
provides important constraints for the investigation of
the times of the previous perihelion, $T_\pi^\ast$, computed
from the solutions based on the orbital arcs whose last
observations were made before the two components of the
fragmented nucleus were optically resolved.  (The first
observation used was, as before, the one made on September~21,
1965.)  Unfortunately, it was possible to compute only
four orbital solutions of this kind (Sekanina \& Kracht
2022), with the last observations on, respectively, October~28,
29, 31, and November~1.  The solutions were based on fewer
than 10~post-perihelion observations.  As already noted,
the fainter nucleus was first detected on November~4.

In spite of the very limited number of post-perihelion
observations available for this category of orbital
solutions, the plot in Figure~3 closely confirms the
results presented by Sekanina \& Kracht (2022).  They
determined that the previous perihelion took place in
\mbox{$1139.9 \pm 2.0$}.  From differently weighted data
points and a simple linear extrapolation I now find
\mbox{$T_0 = 1141.0 \pm 0.6$}.  Both are in good
agreement with the much more reliably determined result
based on the sequence of orbital solutions involving
the principal post-perihelion nucleus.

As seen from Figure 3, the average rate of change in
the time of the previous perihelion, computed from
the unresolved pair of nuclear fragments, equals
\begin{equation}
\left\langle \! \frac{\partial T_\pi^\ast}{\partial
 t_{\rm fin}} \! \right\rangle = -1.88 \; {\rm years
 \; per \; day}
\end{equation}
and satisfies the expected condition
\begin{equation}
\left|\! \left\langle \!\frac{\partial T_\pi}{\partial
 t_{\rm fin}}\! \right\rangle \!\right| \!<\! \left| \!
 \left\langle \!\frac{\partial T_\pi^\ast}{\partial t_{\rm
 fin}} \!\right\rangle \!\right| \!<\! \left| \!\left\langle
 \!\frac{\partial T_\pi^\prime}{\partial t_{\rm fin}}
 \!\right\rangle \!\right| \!.
\end{equation}

In summary, this in-depth orbital analysis of
comet Ikeya-Seki offers a consistent picture
of its history and place in the Kreutz system.
The investigations described in this section
have also been instrumental in identifying the
Chinese comet of September 1138 as the parent
body of Ikeya-Seki\,(Sekanina \& Kracht 2022)
and thereby to settle the stubborn problem of
a ``missing'' second giant Kreutz sungrazer of
the 12th century (Sekanina 2025).

There are both similarities and differences in
comparison to comet C/2026~A1, when it comes
to the issue of the orbital period and its
determination.  The differences are due mainly
to highly uneven sizes of the nuclei of the
two objects.  A similar rationale can be used
to explain differences between comet Ikeya-Seki and
its huge~sibling, whose orbital period is addressed
next.

\section{Comet C/1882 R1 (Great September Comet\\of
 1882)}
Two thirds of Kreutz's (1888, 1891, 1901) monumental
treatise {\it Untersuchungen \"{u}ber das Cometensystem
1843~I, 1880~I und 1882~II\/} dealt with this Great
Comet of 1882.  There was a very good reason for it.
Expressed in simple quantitative terms:\ the 1843 comet
was under scientific observation over a period of
7~weeks and the 1880 comet over two weeks, but the 1882
comet over 37~weeks!

Kreutz sungrazers are known to fade rapidly, and the 1882
comet is in this respect an unrivaled exception, almost
certainly because of its similarly unparalleled perihelion
fragmentation of the nucleus.  Investigating it in
considerable detail, Kreutz (1888, 1891) noted that up
to six secondary nuclei were reported at times.  Four
of them were astrometrically observed often enough that
he was able to determine a separate orbit for each.  He
referred to them as No.~1, \ldots, No.~4 in the order
of their increasing angular distance from the Sun.  In
conformity with the current nomenclature, I assign them
the letters A, B, C, and D, respectively.

Extensive evidence on the split nucleus was a godsend for
the physics of comets because it demonstrated that the
nuclei of the sungrazers were poorly cemented and the
tidal force of the Sun was able to destroy their limited
cohesion.  On the other hand, a multiple nucleus profoundly
complicated orbit-determination efforts, because the number
and configuration of the nuclei varied from day to day and
so did their brightness.  For each of the hundreds of
observations Kreutz had to make a judgment which nuclei
were seen on any given occasion.  And since relative
positions of the nuclei were often hard to determine
because of large uncertainties, he decided to employ
their positions in the direction perpendicular (senkrecht)
to the orbit, because they did line up at an angle with the
direction to the Sun.  A set of elements for each of the
four secondary nuclei was the outcome of Kreutz's meticulous
computations, an effort that must have been astonishingly
time-consuming, given the means of computation available at
the time.

The sheer number of observations of the comet's principal
nucleus~B, extending over a period of about eight months,
and the object's apparently giant dimensions cooperated to
make the result of Kreutz's computations --- its orbit in
general and the orbital period in particular --- exceptionally
accurate, in fact more so than were the orbital periods of
the nuclear components of comet Ikeya-Seki in Marsden's sets
of elements.  An outcome of these unusual circumstances was
that Kreutz's work on the Great Comet of 1882, and specifically
the long orbital period, was the proverbial ``{\it final
straw\/}'' that terminated the protracted controversy on the
bright historical sungrazers as recurring returns of a single
object.  Moreover, in the light of the rather desperate search
for the long-disputed parent of the 1882 and 1965 comets, the
predicted perihelion time --- April 1138 --- that Marsden (1967)
obtained by integrating the same Kreutz orbit, was an {\small
\bf absolutely phenomenal} feat; the estimated peri\-helion time
was missed by four months, or 0.05~percent of the orbital period!
{\small \bf Surprise No.~1}.

%
% Marsden's perfect prediction of the time
% of the previous perihelion of the 1882 comet by integrating
% Kreutz's orbit of nucleus~B back to the 12th century.
% As noted in Section~3, Marsden~obtained April 1138, a result
% that to him may have been disappointing.  However, in the
% light of our recent successful search for the parent
% sungrazer that split into the 1882 and 1965 comets at its
% 12th-century perihelion (Sekanina \& Kracht 2022), Marsden's
% (1967) predicted time turns out to be {\small \bf absolutely
% phenomenal}, deviating from the estimated perihelion time by
% less than four months, or 0.05~percent of the orbital period!
% {\small \bf Surprise No.~1}.
%

%
\begin{table}[b] % Table 1
\vspace{0.5cm}
\hspace{-0.2cm}
\centerline{
\scalebox{1}{
\includegraphics{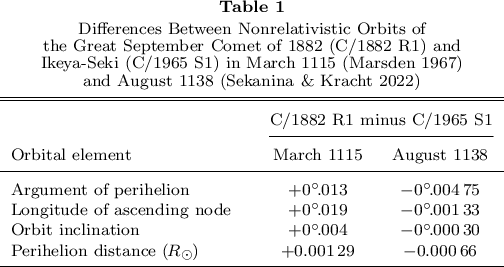}}}
\vspace{0cm}
\end{table}

Next, the other impressive result in Marsden's (1967) paper,
an excellent agreement between the sets of orbital elements
of the 1882 comet and Ikeya-Seki in 1115, could indeed be
judged as virtual proof that the two comets separated from
one another in the early 12th century.~Fol\-lowing the
proposed parenthood of the Chinese comet of September 1138
(Sekanina \& Kracht 2022), the question emerged whether the
orbits of the 1882 and 1965 comets remained at that time
in conformity with each other.  Verification of this point
showed that the agreement in {\small \bf 1138} was in fact
still much {\small \bf better} than in {\small \bf 1115},
as follows from Table~1.  {\small \bf Surprise No.~2}.

\begin{table}[t] % Table 2
\vspace{0.2cm}
\hspace{-0.18cm}
\centerline{
\scalebox{1}{
\includegraphics{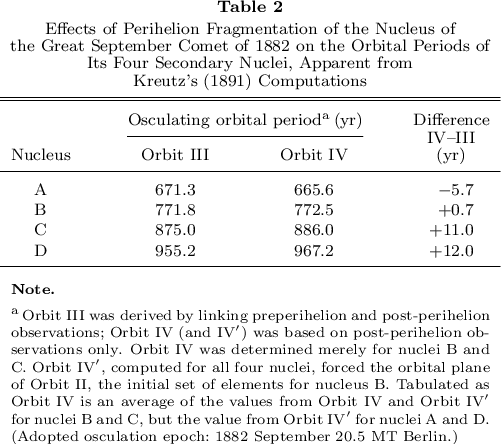}}}
\vspace{0.45cm}
\end{table}

Last but not least, there was a problem of the center of
mass of the 1882 comet after fragmentation.  On the basis
of the distributions of positional residuals from the
orbital solutions that {\small \bf linked preperihelion
and post-perihelion observations} (an {\small \bf
Orbit~III}~category), Kreutz (1891) believed that the center
of mass could~have been nucleus~B or C or D, but not A.
Next to Orbit~III, for each of the four nuclei Kreutz
further computed {\small \bf Orbit~IV}, a set of elements
based on {\small \bf post-perihelion observations} only.
For nuclei~B and C these were general solutions, for all
four nuclei solutions with a forced position of the
orbital plane (thereby not solving for the longitude
of the ascending node and the inclination; runs designated
by Kreutz as {\small \bf Orbits~IV$^\prime$}), a reasonable
approximation.  One should add that Orbit~III was
the same practice used by Marsden (1967) to compute
the orbit of comet Ikeya-Seki (Section~3).  Marsden
did not, and could not, compute Orbit~IV because the
observed post-perihelion orbital arc was much too short
to provide a meaningful set of elements.

It is strange that after completing a great number of
rigorous computations to derive Orbits~III, IV, and
IV$^\prime$, Kreutz summarized his findings in nothing
more than a single, plain statement that it was not
possible to find a better representation of the motions
of the four nuclei than were Orbits~III.  He overlooked
an important~result provided by the orbital periods
of combined Orbits~III and IV, which is presented in
Table~2.  Since the observed preperihelion orbital arc
was much shorter than the post-perihelion arc, its
impact on the computed motion of each nucleus was
limited.  Yet, I demonstrate~in~Table~2 that an
effect, expressed as a difference between Orbit~IV
and Orbit~III in column~4, essentially {\small \bf
disappeared} (within the uncertainty involved)
{\small \bf only for nucleus~B}.  It was this
nucleus that practically {\small \bf coincided
with the comet's center of mass}, and it was its orbital
period~that~fac\-tually~{\small \bf coincided with the
orbital period of the comet before~fragmentation}, equaling
772~years at an oscula\-tion epoch of 1882 September~20
and 744~years de facto. {\small \bf Surprise No.~3}.

It was Orbit~III of nucleus B that Marsden (1967)
integrated back to the 12th century.  The equivalence
of the pre-split orbit with the orbit of nucleus~B,
supported by the data in Table~2, explains Marsden's
remarkably successful prediction of the previous
perihelion in 1138.

The other feature of Table~2 is that for nuclei~A,
C, and D the orbital period of Orbit~III is always
closer to that of the original comet than is Orbit~IV
(or IV$^\prime$).  In other words, the difference
Orbit~IV minus Orbit~III for nucleus~X {\small \bf
correlates} with the difference nucleus~X minus
nucleus~B \mbox{(X = A, C, D)} at least as far
as the sign of the orbital period is concerned.
{\small \bf Surprise No.~4}.

Besides its pedigree value, the distribution of
the orbital periods of the four secondary nuclei
of the Great September Comet of 1882 has other
ramifications, for prognostication in particular.
This issue has recently been dealt with elsewhere
(Sekanina 2026c).

\section{Discussion and Conclusions}
Given that the nucleus of comet C/2026 A1 was much
smaller than the nucleus of comet Ikeya-Seki, which 
was much smaller than the nucleus of the Great Comet
of 1882, the greatly differing problems encountered
with the derived orbital periods in the standard
process of orbit determination are likely to be
nuclear-size dependent.

In the case of comet C/2026 A1 the nucleus was small enough,
definitely subkilometer-sized, so that changes in the pattern
of outgassing over the surface provoked large enough erratic
variations in the momentum of the sublimating gases to trigger
measurable irregular perturbations of the comet's motion
already at heliocentric distances greater than 1~AU.  Because
the outgassing-driven nongravitational acceleration has a
strong tendency to act preferably in the antisolar direction,
the comet's nucleus is effectively moving in a gravity field
that is slightly weaker than the Sun's gravity field.  As a
gravitational solution ignores this effect, the derived
orbital dimensions are larger and the orbital period longer,
as demonstrated for Oort cloud comets with smaller perihelion
distances long ago (Marsden et al.\ 1978).  Nongravitational
solutions account for such effects, but use of any particular
law cannot fit a pattern of outgassing that~is~dom\-inated by
spasmodic variations.  If the rate of outgassing does not
increase along the preperihelion branch of the orbit in
a fairly ordinary fashion, a meaningful orbital period
can only be determined from observations made at rather
large heliocentric distances before perihelion, where
the {\small \bf magnitude of sublimation effects is small}.
Even though the 2026 sungrazer was discovered at a record
distance from the Sun (Maury 2026), the arc of its orbit over
which this condition applied was not long enough to warrant a
solution with the orbital period determined to much better
than about $\pm$15~years (Sekanina \& Kr\'olikowska 2026).

The difficulty with comet Ikeya-Seki was of the same nature
in principle, but it differed greatly in details.  The single
nucleus of the comet approaching the Sun appears to have
been massive enough that the outgassing-driven acceleration
was very probably negligible.  Accordingly, the comet's
preperihelion motion was essentially gravitational and its
integration into the past would have given a fundamentally
correct time of the previous perihelion.  Regrettably, the
comet was discovered only about one month before perihelion
and the observed arc was too short to provide accurate enough
initial conditions for integrating the motion back over
centuries.  The nucleus then split at perihelion into two
components, moving in orbits different from that before
fragmentation.

It was unfortunate that for the principal component this
was not recognized right away, as Marsden (1967) was able to
readily link preperihelion and post-perihelion observations.
The discordance between the motions of the single nucleus
and the principal component was discovered only after an
in-depth examination of the latter orbit as a function of
the last employed observation by Sekanina \& Kracht (2022),
the issue addressed here in Section~3 and Figures~2 and~3.
It is not quite clear to what extent was the discordance an
outcome of the principal fragment entering a new orbit
after the breakup and what was the contribution from a
possible nongravitational acceleration.  In the case of
the fainter fragment the presence of a nongravitational
acceleration was unquestionable.  Yet, in contrast to
{\small \bf C/2026~A1}, whose orbital period exhibited
major inconsistencies a long time {\small \bf before
perihelion}, the more muted problems with comet {\small
\bf Ikeya-Seki} --- a linkage of observations
{\small \bf before} and {\small \bf after perihelion}
in the presence of nuclear fragmentation --- involved
the preperihelion observations only indirectly.

Thanks to the giant sizes of the pre-split nucleus and
the principal secondary nucleus~B of the Great September
Comet of 1882, integration of the gravitational orbit
back over centuries was unusually accurate.  The {\small \bf
orbital period} was subjected to {\small \bf no
measurable change} following nuclear fragmentation at
perihelion, thereby allowing Marsden (1967) to predict
the time of the previous perihelion to within months,
as already noted.

A degree of orbital-period manipulation~was~used~as~a means
to find times of the previous perihelion~of the sungrazers
in Sections~2--3.~The objects that I searched~for were
not on the radar of experts on the Kreutz sungrazer system,
and this was particularly true about the daylight comets
of AD~363.  Before 2021, their potential Kreutz membership had
to my knowledge been brought up only by Seargent (2009) and only
as ``wild speculation.''~It~was after I began to work on the
contact-binary hypothesis~of the Kreutz system (Sekanina 2021)
and searched for a swarm of sungrazers arriving at perihelion
nearly simul\-ta\-neously in the latter part of the 4th century ---
half\-way between Aristotle's comet of 372~BC and~the~Great Comet
of 1106 --- that I came across the reference~to~an account by the
Roman historian Ammianus Marcellinus in Seargent's book, accompanied
by eloquent notes.~That of course was a breakthrough.  And, as if
that~were~not enough, I then registered Marsden's (1967) celebrated
``shot in the dark,'' which was helpful in solving the mystery of
the ``missing'' second 12th-century sungrazer.

The long orbital period of {\small \bf
C\hspace{-0.03cm}/\hspace{-0.01cm}2026~A1 has clinched~the
Kreutz membership of the comets in AD 363}.  Moreover, Sekanina \&
Kr\'olikowska (2026) conclude that if the mass of C/2026~A1
at discovery was a fairly small fraction of the object's mass
at birth in AD~363, the likely parent was {\it Fragment~I\/},
which returned 743~years~later~as the spectacular comet of
1106 and another 737~years~later as the Great March Comet of 1843,
of Population~I.

The orbital periods of the principal components of the split
nuclei of the Great September Comet of 1882 and Ikeya-Seki
were equally instrumental in pointing to the years 1138--1140
as the range of perihelion times for their ``missing'' common
parent, which is now believed to be the Chinese comet of 1138
(Sekanina \& Kracht 2022).  It has been recorded under No.~403
in Ho's (1962) catalogue of ancient and medieval comets and
novae.

There are outstanding questions about the orbital periods
of the two remaining sungrazers mentioned in Section~1, comet
Lovejoy (C/2011~W3) and comet Pereyra (C/1963~R1).  To
pinpoint the parent of comet Lovejoy is rather difficult,
only in part because of the object's headless appearance
from some 1--2~days after perihelion on.  Having derived all
post-perihelion positions of the disintegrated nucleus with
help of an unorthodox technique, Sekanina \& Chodas (2012)
determined a set of elements for a near-perihelion epoch
of osculation.  The orbital period was found to be
\mbox{$698 \!\pm\! 2$ years} and integration of the comet's motion
showed that the previous perihelion would have occurred at
the beginning of 1329, when no bright comet, least of all
a sungrazer, was on record (Ho 1962; Hasegawa 1980).

The disintegration of the nucleus complicated the subsequent
motion of comet Lovejoy because the sunward tip of its
headless condensation was made up of the largest dust
particles that survived intact.  Subjected to nontrivial
solar radiation pressure, the headless comet began to move
in a gravity field that was slightly weaker than that
before the nucleus demise.~The radial, antisolar shift
was taken into account in the orbit-determination routine
and the (essentially constant) normalized magnitude of the
effect of solar radiation pressure was derived to equal
\mbox{$\beta_{\rm tip} = 0.00191 \pm 0.00042$} the solar
gravitational acceleration.
 
One could argue that solar radiation pressure triggered a
perturbation of the radial component of the comet's orbital
acceleration, which was unaccounted for.  Its magnitude at
1~AU from the Sun was $\Gamma$ and its effect, integrated
over times from $t_0$ to $t$, on the true orbital period
$P$ was $\Delta P_{\rm rp}$, for which perturbation theory
gives an expression
\begin{equation}
\frac{\Delta P_{\rm rp}}{P} = \frac{3e \sqrt{p}}{k (1 \!-\!
 e^2)} \, \Gamma \!\! \int_{t_0}^t \!\! g(r) \sin u \, dt,
\end{equation}
where $k$ is the Gaussian gravitational{\vspace{-0.05cm}}
constant (in units of AU$^{\frac{3}{2}}$\,day$^{-1}$), $e$ is
the orbit eccentricity, \mbox{$p = q(1 \!+\! e)$}, $q$ is the
perihelion distance (in AU), $g(r)$ is the perturbation's
dimensionless law of variation with heliocentric distance
\mbox{$r = r(t)$} (in AU) subject to condition \mbox{$g(1) =
1$}, $u$ is a true anomaly at time $t$ (in days).{\vspace{-0.02cm}}
Since the solar gravitational acceleration equals 0.593~cm~s$^{-2}$
at~1~AU from the Sun, the solar radiation pressure acceleration
\mbox{$\Gamma = 2.96 \! \times \! 10^{-4} \beta_{\rm
tip}$}~AU~day$^{-2}$ and \mbox{$g(r) = (r_\oplus/r)^2$},~where
\mbox{$r_\oplus = 1$ AU}.  Next, from Kepler's second law I
substitute in Equation~(11) \mbox{$dt = (r^2/k\sqrt{p}) \, du$},
take a true anomaly at the time of disintegration, $u_{\rm d}$,
to match the lower limit of the integral, $t_0$, and a true
anomaly at aphelion, \mbox{$u = \pi$}, to match the upper
limit, $t$, so that with $\Delta P_{\rm rp}$ and $P$ in the
same units,
\begin{eqnarray}
\Delta P_{\rm rp} & = & 8.88 \! \times \! 10^{-4} \beta_{\rm tip}
 \, P \, \frac{e}{1 \!-\! e^2} \left(\:\!\!\frac{r_\oplus}{k}
 \!\right)^{\!2} \! (1 \!+\! \cos u_{\rm d})  \nonumber \\[0.1cm]
 & \simeq & 1.74 \! \times \! 10^{-5} \beta_{\rm tip}
 \frac{P^{\frac{5}{3}}}{k^2r_{\rm d}},
\end{eqnarray}
where I introduce a heliocentric distance, $r_{\rm d}$, at
the time of disintegration by{\vspace{-0.06cm}} approximating
\mbox{$1 \!+\! \cos u_{\rm d} = p/r_{\rm d}$} in the second
line.  I also use the equality of \mbox{$p/(1 \!-\!e^2) = 
P^{\frac{2}{3}}$} when $P$ is in years, approximate
\mbox{$e = 1$} in the numerator, and leave out $r_\oplus$,
which equals unity.

Inserting now into Equation (12) for comet Lovejoy the above value
of $\beta_{\rm tip}$, \mbox{$P = 683$ years},\footnote{Equivalent
to the osculating orbital period of 698~years.{\vspace{-0.08cm}}}
and \mbox{$r_{\rm d} = 0.144$ AU}, I obtain
\begin{equation}
\Delta P_{\rm rp} = 41 \pm 9 \; {\rm yr}.
\end{equation}
This result suggests that the barycentric orbital period of
comet Lovejoy, corrected for the effect of solar radiation
pressure on the disintegrated headless condensation, equaled
\mbox{683\,--\,41 = 642\,$\pm$\,9 years} and the year of the
previous perihelion passage was, with the same uncertainty,
\mbox{2011\,--\,642 = 1369}, not the start of 1329.~England~(2002)
mentions a possible Kreutz sungrazer seen in Korea and Japan
in the western skies in early March 1368.~It~could be the parent
or previous appearance of comet Lovejoy.\,\,\,\,\,

The orbital motion of comet Pereyra, the last of the five
sungrazers, was investigated by Marsden (1967) and by
Marsden et al.\ (1978).  Possible detection of a secondary
nucleus by Roemer (1963, 1965) on 1963 November~9, nearly
80~days after perihelion, was never confirmed and the comet
in all probability did not split.  Its motion was not integrated
back to the previous perihelion until the problem of the orbital
period was revisited by Sekanina \& Kracht (2022).  They found
that nominally the comet passed perihelion on 1057 April~7, which
indicated an actual orbital period of 906.4~years, but the mean
error was of course $\pm$17~years.  Accordingly, a perihelion
passage between the years 1040 and 1074 would satisfy the
predicted perihelion time to within $\pm$1$\sigma$.

On the strength of evidence provided by two independent lists
of possible historical members of the Kreutz system (Hasegawa
\& Nakano 2001, England 2002), we proposed that the previous
appearance of Pereyra could have been in 1041 (Sekanina \& Kracht
2022), when two bright comets --- both potential Kreutz sungrazers
--- passed perihelion about two months apart.  The actual orbital
period of Pereyra would have been 922~years, longer than the nominal
value by almost 1$\sigma$.  Since the previous perihelion was in
AD~363, the respective orbital period was 678~years and, as a
fragment of one of the comets of 1041, comet Pereyra ended up
in an orbit with a period increased by 244~years.  Even though
such a change is perfectly plausible, one would expect
additional fragments, including one that should have gotten
into an orbit with a period of about 500~years and thus having
arrived at its following perihelion in the course of the
16th century.  Both Hasegawa \& Nakano (2001) and England
(2002) do list possible Kreutz sungrazers in the 16th century,
so that the hypothesis of the relationship between Pereyra
and one of the two comets of 1041 appears to have no
fatal weakness.

Yet, this association is by no means the only possible scenario
for comet Pereyra.  A tendency for the true orbital period to
be shorter than the values derived from observations --- the
trait exhibited by both Ikeya-Seki and C/2026~A1 under very
different circumstances --- makes the relationship of comet
Pereyra with either comet of 1041 somewhat dubious.

Searching for alternative solutions, one obvious possibility
was a candidate object near the time of the nominal predicted
perihelion, April 1057.  England (2002) lists one such suspect,
No.~586 in the {\it Catalogue of Ancient and Naked-Eye Comets\/}
by Hasegawa (1980).  The comet was sighted by Koreans in the
constellation of Corvus during December 1056.  As its historical
record is short, the object apparently was not much of a spectacle
and must have been low above the horizon.

If for whatever reason Marsden's orbital period~of comet Pereyra
should be an overestimate, and its previous appearance in fact
occurred, as in the case of comet Ikeya-Seki, 20 or so years later,
the possible candidate was a Chinese comet with a tail,
seen in the constellation of Scorpius on 1080~January~6, an object
which is listed in Ho's (1962) catalogue under No.~386 and whose
perihelion is estimated by England (2002) to have taken place
four days earlier.

From this narrative it is apparent that among the five Kreutz
sungrazing comets with well established orbital periods, comet
Pereyra is the one with the greatest uncertainty in the time of
the previous appearance, and its motion deserves more attention
in the future.

As final points, it should be emphasized that the documented
high sensitivity of the derived orbital period of a sungrazing
comet to minor perturbations of its motion, prompted by erratic
activity --- such as sudden changes in the pattern of outgassing,
outbursts, or episodes of nuclear fragmentation --- could be
exploited to advantage in cases when the standard approach
does not work or yields inconclusive results.  This is
especially true when observations fail to comply with the
gravitational law, but nongravitational solutions do not
substantially improve the quality of fit or are meaningless.
The propensity of the derived orbital period to effects of
this kind disappears only among very massive sungrazers,
such as the Great September Comet of 1882, because they are
not subjected to measurable nongravitational forces.

Experience gained from the experiments with the few objects
of the Kreutz system suggests that astrometric observations
at the very end of the used orbital arc may have sizable
impact on the value of the derived orbital period.  To an
extent, such data could be used to monitor the timing of
minor perturbations of the orbital motion of a sungrazer,
triggered by the complex physical processes on its nucleus.
However, the ultimate goal behind the efforts to determine
the orbital period of a sungrazing comet as accurately as
possible is dictated by the quest to properly predict the
time of the object's previous perihelion passage, which
touches upon a broad range of issues associated with each
sungrazer's pedigree.

\vspace{0.2cm}
\begin{center}
{\footnotesize REFERENCES}
\end{center}

\vspace{-0.02cm}
\parbox{8.33cm}{\footnotesize
\hspace*{-0.41cm}
England, K.\ J.\ 2002, J.\ Brit.\ Astron.\ Assoc., 112, 13 \\[0cm]
\hspace*{-0.41cm}
Hasegawa, I.\ 1980, Vistas Astron., 24, 59 \\[0cm]
\hspace*{-0.39cm}
Hasegawa, I., \& Nakano, S.\ 2001, Publ.\ Astron.\ Soc.\
 Japan,~53,~931 \\[0cm]
\hspace*{-0.41cm}
Ho, P.-Y.\ 1962, Vistas Astron., 5, 127 \\[0cm]
\hspace*{-0.41cm}
Kreutz, H.\ 1888, Publ.\ Sternw.\ Kiel, No.\ 3 \\[0cm]
\hspace*{-0.41cm}
Kreutz, H.\ 1891, Publ.\ Sternw.\ Kiel, No.\ 6 \\[0cm]
\hspace*{-0.41cm}
Kreutz, H.\ 1901, Astron.\ Abhandl., 1, 1 \\[0cm]
\hspace*{-0.41cm}
Marsden, B.\ G.\ 1967, Astron.\ J., 72, 1170 \\[0cm]
\hspace*{-0.38cm}
Marsden,\,B.\,G., Sekanina,\,Z., \& Everhart,\,E. 1978,
 Astron.\,J.,~83,~64 \\[0cm]
\hspace*{-0.41cm}
Maury, A.\ 2026, Centr.\ Bur.\ Electr.\ Tel.\ No.\ 5658 \\[0cm]
\hspace*{-0.41cm}
Pohn, H.\ 1965, IAU Circ.\ 1937 \\[0cm]
\hspace*{-0.41cm}
Roemer, E.\ 1963, Publ.\ Astron.\ Soc.\ Pacific, 75, 535 \\[0cm]
\hspace*{-0.41cm}
Roemer, E.\ 1965, Astron.\ J., 70, 397 \\[0cm]
\hspace*{-0.41cm}
Seargent, D. 2009, The Greatest Comets in History:\ Broom Stars and
 {\hspace*{0cm}}Celestial Scimitars. Springer Science+Business
 Media, LLC, 260pp \\[0cm]
\hspace*{-0.41cm}
Sekanina, Z.\ 2021, eprint arXiv:2109.01297 \\[0cm]
\hspace*{-0.41cm}
Sekanina, Z.\ 2022, eprint arXiv:2202.01164 \\[0cm]
\hspace*{-0.41cm}
Sekanina, Z.\ 2025, eprint arXiv:2505.14662 \\[0cm]
\hspace*{-0.41cm}
Sekanina, Z.\ 2026a, Centr.\ Bur.\ Electr.\ Tel.\ No.\ 5681 \\[0cm]
\hspace*{-0.41cm}
Sekanina, Z.\ 2026b, eprint arXiv:2602.17626 \\[0cm]
\hspace*{-0.41cm}
Sekanina, Z.\ 2026c, eprint arXiv:2605.09938 \\[0cm]
\hspace*{-0.39cm}
Sekanina, Z., \& Chodas, P.\ W.\ 2012, Astrophys.\ J., 757, 127 (33pp)\\[0cm]
\hspace*{-0.41cm}
Sekanina, Z., \& Kracht, R.\ 2022, eprint arXiv:2206.10827 \\[0cm]
\hspace*{-0.41cm}
Sekanina, Z., \& Kr\'olikowska, M.\ 2026, eprint arXiv:2607.25939}
\vspace{0.4cm}
\end{document}